\documentclass[prb,twocolumn]{revtex4}

\usepackage{times}
\usepackage{amsmath}
\usepackage{amssymb}
\usepackage{psfrag}
\usepackage{pstricks}
\usepackage{graphicx}% Include figure files
\usepackage{dcolumn}
\usepackage{bm}
\usepackage[hypertex]{hyperref}

\def\be{\begin{equation}}
\def\ee{\end{equation}}
\def\nn{\nonumber}

\begin{document}

\title{Edge States and Interferometers in the Pfaffian and anti-Pfaffian States}

\author{Waheb Bishara}
\affiliation{ Department of Physics, California Institute of
  Technology, MC 256-80 Pasadena, CA 91125}
\author{Chetan Nayak}
\affiliation{Microsoft Research, Station Q, CNSI Building,
University of California, Santa Barbara, CA 93106-4030}
\affiliation{Department of Physics,
University of California, Santa Barbara, CA 93106}

\begin{abstract}
We compute the tunneling current in a double point contact geometry of a Quantum Hall system at filling fraction $\nu=5/2$, as function of voltage and temeprature, in the weak tunneling regime. We quantitatively compare two possible candidates for the state at $\nu=5/2$: the Moore-Read Pfaffian state, and its particle-hole conjugate, the anti-Pfaffian. We find that both possibilities exhibit the same qualitative behavior, and both have an even-odd effect that reflects their non-Abelian nature, but differ quantitatively in their voltage and temperature dependance.
\end{abstract}
\maketitle

\section{Introduction}

Quantum Hall (QH) devices at certain filling fractions are the only systems known to be in topological phases. The $\nu=1/3$ Laughlin state is in an Abelian topological phase. The excitations of such a phase carry a fraction of an electron charge and have fractional statistics which are intermediate between bosonic and fermionic statistics. The fractional charge has been confirmed experimentally \cite{Goldman95,Picciotto97,Saminadayar97},
and experiments showing indications of fractional statistics have
been recently performed \cite{Camino05}.

The observed\cite{Willett87,Eisenstein02,Xia04} Quantum Hall state at filling fraction $\nu=5/2$ is the primary candidate for a system in a non-Abelian topological phase, and is believed to be described by the Moore-Read Pfaffian state \cite{Moore91,Greiter92}
as a result of numerical evidence \cite{Morf98,Rezayi00}.
The excitations of the Pfaffian carry fractional charge $e/4$ and have
non-Abelian braiding statistics: for given quasiparticle positions, there are several linearly-independent quantum states of the system, and braiding the quasiparticles causes a rotation in this
space\cite{Nayak96c,Read96,Fradkin98,Read00,Ivanov01}.
In addition to their novelty, these properties could be useful for topological quantum computation \cite{DasSarma05}.

In the absence of Landau Level mixing, the Hamiltonian of a half-filled Landau level is particle-hole symmetric. The Pfaffian state, if it is the ground state of such
a Hamiltonian, spontaneously breaks particles-hole symmetry. The particle-hole
conjugate of the Pfaffian, dubbed the anti-Pfaffian \cite{LeeSS07,Levin07},
has exactly the same energy as the Pfaffian in the absence of Landau level mixing.
Hence, it is a serious candidate for the $\nu=5/2$ state observed in experiments,
where Landau level mixing, which is not small, will favor one of the two states.
Therefore, it is important to find experimental probes which can
distinguish between these two states. Although the two states are
related by a particle-hole transformation and are both non-Abelian,
they differ in important ways: their quasiparticle statistics
differ by Abelian phases, and the anti-Pfaffian has three
counter-propagating
neutral edge modes while the Pfaffian edge is completely chiral.
In this paper we consider edge tunneling experiments for both the Pfaffian and the anti-Pfaffian states, and we find quantitative differences between the two resulting
from these distinctions.

\begin{figure}
  \includegraphics[width=3in]{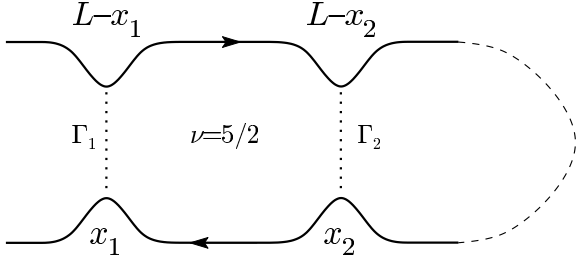}
  \caption{The double point contact geometry. Edge quasiparticles can tunnel between the top and bottom edges at the point contacts $j=1,2$, with tunneling amplitude $\Gamma_j$. The dashed line serves as a reminder that both top and bottom edges are two sections of the same edge. The two ends of point contact $j$ are two points on the same edge separated by a distance $L-2x_j$, where $L$ is the length of the edge.}\label{2PC}
\end{figure}

The double point contact geometry has been proposed as a probe for non-Abelian statistics \cite{Fradkin98,Bonderson06a,Stern06,Ardonne07b,Fidkowski07,
Overbosch07,Rosenow07b}. In this setup, a QH bar is gated so that two constrictions are created, as shown in Figure \ref{2PC}, and quasiparticles can tunnel from one edge to the other at either constriction. The dashed line in Fig. \ref{2PC} serves as a reminder that
the two edges are actually different sections of a single edge which
is the boundary of the system; consequently, inter-edge tunneling satisfies
topological conservation laws which are important in the non-Abelian case.
An edge quasiparticle entering the sample from the left can tunnel to the lower edge through either point contact, and the measured tunneling current is sensitive to the interference between these two possible trajectories. The phase difference between the quantum amplitudes of these two trajectories depends on the applied voltage between the top and bottom edges, the magnetic flux enclosed between the two trajectories, and the number of quasiparticles localized in the bulk between the two trajectories.
If the quasiparticles have non-Abelian statistics,
the quantum state of the system can change when the edge
quasiparticle encircles the localized bulk ones, and the effect on the interference
term is more than merely a phase shift. The Pfaffian and anti-Pfaffian states
exemplify the most extreme case: if there is an even number of localized quasiparticles enclosed between the tunneling trajectories, there will be interference that depends on the magnetic flux and applied voltage, while in the presence of an odd number of bulk quasiparticles in the bulk, the interference pattern will be completely lost. We will recover these striking results using an explicit edge theory calculation.

The visibility of the interference pattern in the even quasiparticle
case will be obscured by thermal smearing as well as
the difference between the charged and neutral mode
velocities. Naively, the latter is particularly acute in the anti-Pfaffian
case, where the velocities have opposite sign. However, as we will
see quantitatively from the edge state calculation below,
the difference between the even and odd quasiparticle cases
will be visible for sufficiently low temperature in both the Pfaffian
and anti-Pfaffian states. The required temperature
vanishes as the distance between the contacts or the difference in
velocities is increased.

The principle conceptual difficulty
in analyzing inter-edge tunneling stems from the non-Abelian
nature of the bulk state, which causes ambiguities in edge correlation
functions (or, more properly, {\it conformal blocks}). We show how
these are resolved, following
Refs. \onlinecite{Fendley06,Fendley07a} and further refinements
introduced in Refs. \onlinecite{Fidkowski07,Ardonne07b}.

The paper is organized as follows. In Section \ref{main} we set up the perturbative calculation to lowest order, explain the ambiguity that arises in evaluating correlation functions due to the non-Abelian nature of the edge, and show how to resolve this ambiguity,
following Refs. \onlinecite{Fendley06,Fendley07a}. In Section \ref{TV}, we find the expected tunneling current behavior as a function of bias voltage and temperature in the Pfaffian state, taking into consideration the different velocities of charged and neutral modes on the edge.
We show that for sufficiently low temperature, interference will be visible
in the even quasiparticle case. In Section \ref{aPf},
we repeat the calculation for the anti-Pfaffian state and show the quantitative differences with the Pfaffian case.

\section{Tunneling Operators and Conformal Blocks}\label{main}

We now set up the calculation of the tunneling current to lowest order
and discuss the basic issues which arise. The Pfaffian and anti-Pfaffian
cases are conceptually similar, so we focus on the Pfaffian for the sake of
concreteness.
The edge theory of the Pfaffian state has a chiral bosonic charge mode and a chiral neutral Majorana mode\cite{Milovanovic96,Bena06,Fendley06,Fendley07a}
\be
\label{eqn:Pf-edge}
{\cal L}_{\rm Pf}(\psi,\phi)=\frac{2}{4\pi}\partial_x\phi\left(\partial_t+v_c\partial_x\right)\phi + i\psi\left(\partial_t+v_n\partial_x\right)\psi
\ee
Both modes propagate in the same direction, but will have different velocities in general. One expects the charge velocity $v_c$ to be larger than the neutral velocity $v_n$. The electron operator is a charge $1$ fermionic operator:
\be
\Phi_{el}=\psi e^{i\sqrt{2} \phi}
\ee
and the $e/4$ quasiparticle operator is:
\be
\Phi_{1/4}=\sigma e^{i\phi/2\sqrt{2}}
\ee
where $\sigma$ is the Ising spin field of the Majorana fermion
theory\cite{Fendley07a}. When inter-edge
tunneling is weak, we expect the amplitude $\Gamma$ for charge-$e/4$ to be transferred from one edge to the other to be larger than for higher charges
$ne/4$, which should be $\sim\Gamma^n$. It is also the most relevant tunneling
operator in the Renormalization Group sense \cite{Fendley06,Fendley07a},
so we will focus on it. Since it is relevant, its effective value grows
as the temperature is decreased, eventually leaving the weak
tunneling regime. We assume that the temperature is
high enough that the system is still in the weak tunneling regime
and a perturbative calculation will be valid, but still much lower than
the bulk energy gap.

Following Ref. \onlinecite{Chamon97}, we write the tunneling Hamiltonian in the
form:
\begin{eqnarray}
\label{eqn:tunneling-Ham}
H_\text{t}(t) &=& \Gamma_1 e^{-i{\omega^{}_J}t}\, V_1(t) +
\Gamma_2 \,e^{i\Phi/4\Phi_0}\: e^{-i{\omega^{}_J}t}\,V_2(t)\cr
& & +\: \text{h.c.}
\end{eqnarray}
The frequency $\omega^{}_J=\frac{eV}{4}$ is the Josephson frequency for a charge $e/4$
quasiparticle with voltage $V$ applied between the top and bottom edges.
The difference in the magnetic fluxes enclosed by the two
trajectories around the interferometer is $\Phi$. We have chosen a gauge in which
the vector potential is concentrated at the second point contact so that $\Phi$
enters only through the second term above.
Both edges are part of the boundary of the same
Hall droplet, so we can denote the point on the upper edge
which is on the other side of point contact $j$
from $x_j$ by $L-{x_j}$, where $j=1,2$ and $L$ is large.
The operator ${V_j}(t)$ tunnels a quasiparticle between $x_j$ and $L-x_j$:
\begin{align}
V_j(t)=\sigma(x_j,t)\sigma(L-x_j,t)e^{\frac{i}{\sqrt{8}}\phi(x_j,t)}e^{-\frac{i}{\sqrt{8}}\phi(L-x_j,t)}
\end{align}
The current operator can be easily found from the commutator of the tunneling
Hamiltonian and the charge on one edge:
\begin{multline}
\label{eqn:current-op}
I(t)=\frac{ie}{4}\left(\Gamma_1 e^{-i\omega^{}_J t}V_1(t) -
\text{h.c.}\right)\\
+ \frac{ie}{4}\left(\Gamma_2 \,e^{i\Phi/4\Phi_0}\: e^{-i{\omega^{}_J}t}\,V_2(t)
- \text{h.c.}\right)
\end{multline}
To lowest order in perturbation theory, the tunneling current is found to be:
\be\label{Current}
\langle I(t) \rangle = -i \int_{-\infty}^t dt' \, \langle 0 | [I(t),H_\text{t}(t')]|0\rangle
\ee
In order to compute the current, we substitute (\ref{eqn:tunneling-Ham})
and (\ref{eqn:current-op}) into (\ref{Current}). We obtain:
\begin{multline}
I(t) = \frac{e}{4}\sum\limits_{j,k}
{\Gamma_j}{\Gamma_k^*}\: e^{i(j-k)\Phi/4\Phi_0}\:\times \\
\int_{-\infty}^t dt' \, e^{i{\omega^{}_J} (t'-t)} \left(
\langle {V_j}(t)  {V_k^\dagger}(t')\rangle -  \langle {V_k^\dagger}(t')
{V_j}(t) \rangle
\right)
\end{multline}

Therefore, we must compute the correlation function
\begin{align}\label{VjVk}
\langle V_j(t) & V_k^\dagger(t')\rangle=  \\
 &\langle \sigma(x_j,t)\sigma(L-x_j,t)\sigma(L-x_k,t')\sigma(x_k,t')\rangle \times \nn \\
 &\langle e^{\frac{i}{\sqrt{8}}\phi(x_j,t)}e^{-\frac{i}{\sqrt{8}}\phi(L-x_j,t)}e^{\frac{i}{\sqrt{8}}\phi(L-x_k,t')}e^{-\frac{i}{\sqrt{8}}\phi(L-x_k,t')}\rangle \nn
\end{align}
This correlation function is at the heart of our calculation. The correlations involving the bosonic fields are straightforward to calculate and, in the limit of a long sample, $L\rightarrow \infty$, the bosonic correlation function
breaks into a product of two-point correlation functions of fields
on the same edge:
\begin{align}
 &\langle e^{\frac{i}{\sqrt{8}}\phi(x_j,t)}e^{-\frac{i}{\sqrt{8}}\phi(x_k,t')}\rangle\langle e^{-\frac{i}{\sqrt{8}}\phi(L-x_j,t)}e^{\frac{i}{\sqrt{8}}\phi(L-x_k,t')}\rangle \nn \\
 &= \prod_{r=\pm}\left[ \delta+i\left(v_c(t-t')+r(x_j-x_k)\right)\right]^{-1/8}
\end{align}
However, the four $\sigma$ correlation function is actually ill-defined without further information, namely the fusion channels of the four $\sigma$ operators.
(Technically, the correlation function is what is called a {\it conformal block}.)
These are determined by the physical situation, as we elaborate on this below.

In the Ising Conformal Field Theory, the $\sigma$ operators have non-trivial fusion rules:
\be\label{sigmasigma}
\sigma \times \sigma = I + \psi
\ee
A correlation function of $2n$ $\sigma$ particles is non-vanishing only if all
of the operators fuse together to the identity, but there are a number of ways in which the fields can do that. In the four $\sigma$ operators case, the correlation $\langle\sigma(z_1)\sigma(z_2)\sigma(z_3)\sigma(z_4)\rangle$ has two different conformal blocks corresponding to the two possible fusions. In the standard
notation explained, for instance, in this context in
Ref. \onlinecite{Fendley07a}, these two conformal blocks/fusion channels are:
\begin{center}
\begin{picture}(140,30)
\put(-20,3){${\cal F}_c\equiv$}
\put(13,3){$I$}
\put(30,20){$1$}
\put(50,20){$2$}
\put(90,20){$3$}
\put(110,20){$4$}
\put(70,3){$c$}
\put(10,0){\line(1,0){125}}
\put(30,0){\line(0,1){15}}
\put(50,0){\line(0,1){15}}
\put(90,0){\line(0,1){15}}
\put(110,0){\line(0,1){15}}
\put(125,3){$I$}
\end{picture}
\end{center}
where $c=1$ or $\psi$ is the fusion product of the fields at the
space-time points $z_1$ and $z_2$. Their explicit forms are:
\begin{eqnarray} \label{blocks}
\nonumber
{\cal F}_I &=& \left(\frac{1}{z_{12}z_{34}(1-x)}\right)^{1/8}
\left(1+\sqrt{1-x}\right)^{1/2}, \\
{\cal F}_\psi &=& \left(\frac{1}{z_{12}z_{34}(1-x)}\right)^{1/8}
\left(1-\sqrt{1-x}\right)^{1/2},
\label{fourpointexplicit}
\end{eqnarray}
where $z_{ij}=z_i-z_j$ and $x=z_{12}z_{34}/z_{13}z_{24}$.

\begin{figure}
  \includegraphics[width=3.5in]{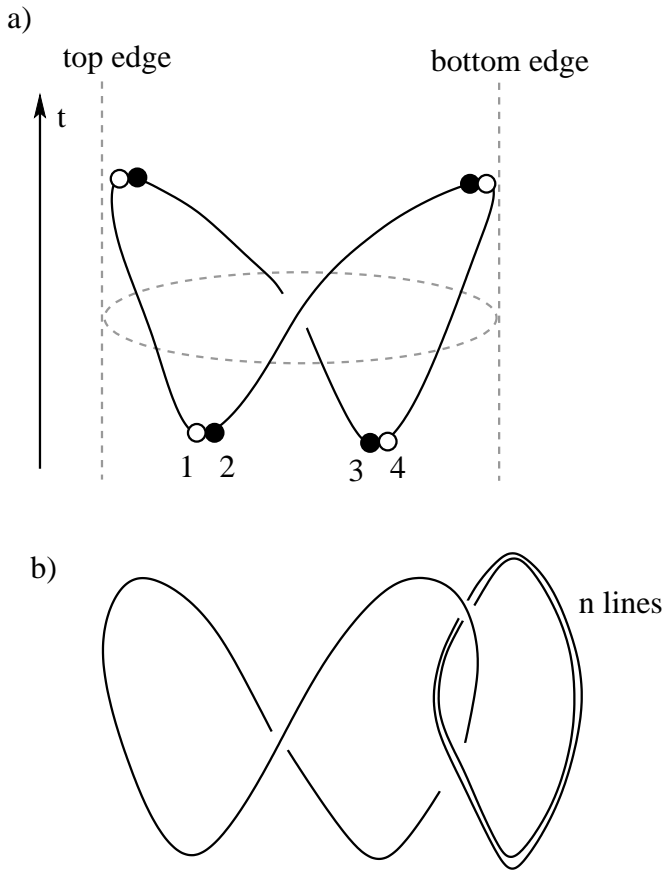}\\
  \caption{(a) The knot corresponding to the matrix element between the state
  in which two quasiparticle-quasihole pairs, $1$, $2$ and $3$ $4$ are created out of the vacuum (top half)
  and go to opposite edges and and a state in which the two quasiparticles on each edge (e.g. $1$, $3$
  on the top edge) fuse to $I$.
  Equivalently, it is one element of the $F$-matrix, which transforms between the basis
  of conformal blocks in which $1$, $2$ has a fixed fusion channel (and, therefore, $3$, $4$ does as well)
  and the basis in which $1$, $3$ has a fixed fusion channel.
    (b) The same matrix element with $n$ quasiparticles in the bulk. The $n$ quasiparticles
    are assumed to have been created in pairs in the distant past, with one member of each pair
    taken inside the interferometer and the other member left outside. The two tunneling events
    are assumed to occur at different point contacts. The figure then gives the matrix element
    between the states in which $1$, $2$ and $3$, $4$ are created out of the vacuum (top half),
   go to opposite edges, and encircle the bulk qusiparticles; and a state in which $1$, $3$
   fuse to $I$.}\label{braids}
\end{figure}

Now for an obvious question: which conformal block enters the perturbative
calculation?  As explained in Ref. \onlinecite{Fendley06,Fendley07a}, when
there are no quasiparticles in the bulk, the correct
choice is the conformal block in which the $\sigma$ operators in the tunneling operator $V_j(t)$, i.e. $\sigma(x_j,t)$ and $\sigma(L-x_j,t)$, fuse to the identity. Since all this operator does is transfer a quasiparticle from one side of the Hall sample to the other,
it should not change the topological charge on the edge, which
would involve the creation of a fermion.
In the bottom half of Fig. \ref{braids}a, we show two successive
tunneling events. Each can be envisioned as the creation out of the vacuum
of a quasiparticle-quasihole pair in the bulk. Saying that they are created `out of the vacuum'
is equivalent to saying that they fuse to $I$.
The quasiparticle then goes to one edge and the quasihole goes to the other.
A second tunneling event (either at the same or a different
point contact) occurs in the same way. Let us, for the sake of
concreteness call the quasiparticle and quasihole which are created in the first tunneling process
$1$ and $2$; in the second tunneling process, $3$ and $4$ are created.
(For these purposes, there is no need to distinguish between quasiparticles
and quasiholes.) Let us assume that quasiparticles $1$ and $3$ go to the top edge
while $2$ and $4$ go to the bottom edge.
If the two edges are independent
(as occurs in the $L\rightarrow\infty$ limit),
this process has a non-zero amplitude only if $1$, $3$ fuse to $I$ and $2$, $4$ fuse
to $I$, as depicted in the top half
of Fig. \ref{braids}a. ($I$ is depicted by the absence of a line. If a fermion
were the result of fusing the two quasiparticles, there would be a wavy line
emanating upward from each of the two fusion points at the top of Fig. \ref{braids}a.)
This picture can be interpreted as the matrix element
between the state in which quasiparticle-quasihole pairs $1$, $2$ and $3$, $4$
are created in the bulk and go to the edges (bottom) and the state in which quasiparticles
$1$, $3$ fuse to $I$ and $2$, $4$ fuse to $I$ (top).

Hence the correlation function in Eq.(\ref{VjVk}) is
actually the conformal block:
\begin{center}
\begin{picture}(210,30)
\put(-20,3){${\cal F}_I = $}
\put(13,3){$I$}
\put(15,20){$(x_j,t)$}
\put(45,20){$(L-x_j,t)$}
\put(100,20){$(L-x_k,t')$}
\put(150,20){$(x_k,t')$}
\put(90,3){$I$}
\put(10,0){\line(1,0){180}}
\put(30,0){\line(0,1){15}}
\put(55,0){\line(0,1){15}}
\put(135,0){\line(0,1){15}}
\put(165,0){\line(0,1){15}}
\put(180,3){$I$}
\end{picture}
\end{center}

On the other hand, in the $L\rightarrow \infty$ limit, we expect the $\sigma$ correlation in Eq.(\ref{VjVk}) to break into a product of correlators of fields on
the same side of the sample:
\be
\langle \sigma(x_j,t)\sigma(x_k,t')\rangle \langle \sigma(L-x_j,t)\sigma(L-x_k,t')\rangle
\ee
As noted above, this correlation function is non-vanishing only if the fields on the same side of the sample fuse to the identity. This conformal block is given pictorially by:
\begin{center}
\begin{picture}(210,30)
\put(-20,3){${\cal G}_I = $}
\put(13,3){$I$}
\put(15,20){$(x_j,t)$}
\put(45,20){$(x_k,t')$}
\put(100,20){$(L-x_j,t)$}
\put(150,20){$(L-x_k,t')$}
\put(90,3){$I$}
\put(10,0){\line(1,0){180}}
\put(30,0){\line(0,1){15}}
\put(55,0){\line(0,1){15}}
\put(135,0){\line(0,1){15}}
\put(165,0){\line(0,1){15}}
\put(180,3){$I$}
\end{picture}
\end{center}
In the ${\cal G}_c$s, we specify the fusions of fields on the same side of the edge rather than
opposite sides of a point contact. In the $L\rightarrow\infty$ limit, ${\cal G}_\psi$ vanishes. The conformal blocks ${\cal G}_c$ are linear combinations of the
${\cal F}_c$s; both form bases for the two-dimensional vector
space of conformal blocks. The basis change between the
two is called the $F$-matrix, which is part of the basic data
characterizing a topological phase. We can write:
\be \label{basischange}
{\cal F}_I=a_I {\cal G}_I+a_\psi {\cal G}_\psi
\ee
where the coefficients $a_I$ and $a_\psi$ are two of the entries
in the $F$-matrix. They can be calculated by computing the Kauffman bracket
for a braid that corresponds to this
change of basis, as was done in Ref. \onlinecite{Fendley06,Fendley07a}:
\begin{equation}
{a_I} = \frac{1}{\sqrt{2}} \hskip 0.4 cm \mbox{(no qps in bulk)}
\end{equation}
For the purposes of our calculation, we only need
the long sample limit of the correlation function ${\cal F}_I$.
As explained above, we find that it is proportional to ${\cal G}_I$, which can be easily evaluated since it is simply the product of two two-point correlation functions (it can also be obtained by taking the large-$L$ limit of the expression for $G_I$
as in Eq.(\ref{blocks})):
\begin{align} \label{LongSample}
{\cal F}_I \mid_{L\rightarrow \infty} &= a_I {\cal G}_I\mid_{L\rightarrow_\infty} \\
&= a_I\prod_{\epsilon=\pm}\left[ \delta+i\left(v_n(t-t')+\epsilon(x_j-x_k)\right)\right]^{-1/8} \nn
\end{align}

We now generalize this to the case in which there are
$n$ quasiparticles in the bulk between the two point contacts.
Correlation functions in which all of the fields are at the same
point contact are unchanged. However, as pointed out in Refs. \onlinecite{Ardonne07b,Fidkowski07},
when two tunneling processes occur at different point contacts, the two quasiparticle-quasihole pairs
are created out of the vacuum as before, but quasiparticle $1$
must encircle the bulk quasiparticles before it can be fused with quasiparticle
$3$. This difference modifies the matrix element with the state in which $1$, $3$ fuse
to $I$ and $2$, $4$ fuse to $I$, as depicted in Figure \ref{braids}b. Let us consider the simplest case, in which
there is a single quasiparticle in the bulk. We can imagine that a quasiparticle-quasihole pair was
created in the distant past and one member of the pair was brought into the interferometer
while the other member was left outside. Then we create the quasiparticle-quasihole pairs
$1$, $2$ and $3$, $4$ and take $4$ around the bulk quasiparticle. This process is depicted
in the bottom half of Fig. \ref{braids}b.
We can compute the resulting $a_I$ by computing the matrix element between the resulting
state and the state in which $1$, $3$ fuse to $I$ (as do $2$, $4$). This matrix element
can be computed from the Kauffman bracket of the link in Fig. \ref{braids}b or, equivalently,
by using the $F$ and $R$ matrices of the theory. By either method, we find
${a_I}=0$. The reason is that, after $4$ is taken around the bulk quasiparticle, either $1$, $3$ or
$2$, $4$ (but not both) must fuse to $\psi$ rather than $I$. Therefore, there is no amplitude
for $1$, $3$ and $2$, $4$ to fuse to $I$. For the same reason, ${a_\psi}=0$, so even for
$L$ finite, there is no contribution from such a process. In fact, the same result is obtained
for any odd $n$ since an odd number of quasiparticles must fuse to $\sigma$. Therefore,
their effect is the same as if there were a single quasiparticle in the bulk:
\begin{equation}
{a_I}= 0 \hskip 0.4 cm \mbox{(odd number $n$ of qps in bulk)}
\end{equation}
For $n$ even, the $n$ bulk quasiparticles can fuse to
either $I$ or $\psi$. The former case is the same as in the absence of quasiparticles;
in the latter case, there is an additional minus sign which is acquired when a $\sigma$
goes around a $\psi$:
\begin{equation}
{a_I}=\pm \frac{1}{\sqrt{2}}  \hskip 0.4 cm \mbox{(even number $n$ of qps in bulk)}
\end{equation}

With the correct conformal block in hand, as specified by the corresponding
value of $a_I$, we can now give a meaning to expressions such as
(\ref{VjVk}) and can use (\ref{Current}) to compute the current through our interferometer.

In the preceding discussion, we have focussed on the neutral sector of the theory,
where the interesting non-Abelian effects occur. However, there is also a charged
sector of the theory. The full conformal theory describing the edge includes both
parts. As a result, there are additional phases which result from the change of
basis when there are quasiparticles in the bulk. Furthermore, we must exercise a little
more care in distinguishing quasiparticles from quasiholes since they have different
Abelian phases. By recalculating Fig. \ref{braids}b
with the Abelian part of the theory included, we find that $a_I$ acquires
an additional phase $n\pi/4$ when there are $n$ quasiparticles in the bulk
and $1$ and $4$ are quasiparticles while $2$ and $3$ are quasiholes.
The opposite phase results when $2$ and $3$ are quasiparticles
while $1$ and $4$ are quasiholes.

\section{Temperature And Voltage Dependence
of the Current through an Interferometer} \label{TV}

To lowest order in $\Gamma_1$, $\Gamma_2$, the current
naturally breaks into the sum of three terms:
\begin{equation}
I = {I_1} + {I_2} + I_{12}
\end{equation}
where
\begin{equation}
{I_j} = \frac{e}{4}
\left|\Gamma_j\right|^2 \,
\int_{-\infty}^0 dt \, e^{i{\omega^{}_J}t} \left(
\langle {V_j^{}}(0)  {V_j^\dagger}(t)\rangle -  \langle{V_j^\dagger}(t)
{V_j^{}}(0) \rangle
\right)
\end{equation}
and
\begin{eqnarray}
I_{12} &=& \frac{e}{4}
\Gamma_1\Gamma_2^*\,
\int_{-\infty}^0 dt \, e^{i{\omega^{}_J} t} \left(
\langle {V_1^{}}(0)  {V_2^\dagger}(t)\rangle -  \langle {V_2^\dagger}(t)
{V_1^{}}(0) \rangle
\right)\cr
& & \: \: \:\: + \:\: \text{c. c.}
\end{eqnarray}
$I_{j}$, $j=1,2$ would be the backscattered current
if only point contact $j$ were present.
$I_{12}$ is
due to interference between the process in which
a quasiparticle tunnels between the two edges
at $x_1$ and the process in which it continues to $x_2$ and tunnels there. As a result, $I_{12}$ depends on the magnetic flux and
the number of bulk quasiparticles between the two point contacts;
it reflects the non-Abelian statistics of quasiparticles, namely the
difference between even and odd numbers of bulk quasiparticles.
Meanwhile, ${I_1}, {I_2}$, and $I_{12}$ all depend on the bias voltage and
temperature. In this section we quantitatively analyze the
dependence of $I$ on all of these parameters.

\begin{figure}
  \includegraphics[width=3in]{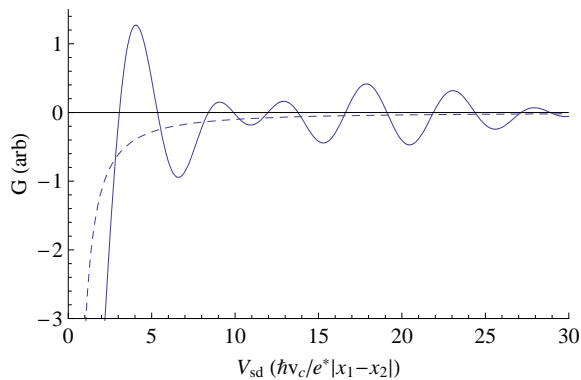}
  \caption{The differential conductance as a function of applied voltage at low temperature, for $v_n=0.75v_c$. The dashed line is the conductance with an odd number of quasiparticles in the interference loop, and the solid line is for an even number. The charge and neutral velocities can be extracted from the the two oscillation periods.}\label{Conductance}
\end{figure}

We first consider the zero-temperature case.
The single point contact current term, $I_{1}+I_{2}$,
is identical to the backscattering current due a single impurity in a
Luttinger Liquid.
The current is a power law in voltage:
\begin{multline}
\label{eqn:zero-temp-direct}
I_{1}+I_{2}=\frac{1}{\sqrt{2}}  \, \frac{e}{4} \, \frac{\pi}{\Gamma (\frac{1}{2})}\left(|\Gamma_1^2|+|\Gamma_2|^2\right)v_n^{-1/4} v_c^{-2/8}\,
\times\\
\text{sgn}(V) \left(\frac{e|V|}{4}\right)^{-1/2}
\end{multline}
The factor of $\frac{1}{\sqrt{2}}$ is $a_I$ discussed in the previous section.
We now consider $I_{12}$. For an odd number of quasiparticles in the bulk,
\begin{equation}
I_{12} = 0 \hskip 0.6 cm \mbox{(odd number $n$ of qps in bulk)}
\end{equation}
For an even number $n$ of quasiparticles in the bulk,
$I_{12}$ can be evaluated analytically in the special case in which
the two velocities are equal:
\begin{multline}
\label{eqn:zero-temp-int}
I_{12} =\pm \frac{1}{\sqrt{2}} \, \frac{e}{4} \, \frac{\pi^{3/2}2^{9/4}}{\Gamma(\frac{1}{4})} |\Gamma_1||\Gamma_2|
 \cos\!\left(\frac{\Phi}{4\Phi_0}+
n\frac{\pi}{4}+\alpha\right)\,\times\\
 \text{sgn}(V) \, |V|^{-1/2}\,\times\\
  \,\left(\frac{e|x_1-x_2|}{4 v}|V|\right)^{\!1/4}\,
\!  J_{-1/4}\!\left(\frac{e|x_1-x_2|}{4v}
 |V|\right)
\end{multline}
In this expression, the $\pm$ sign is obtained if the quasiparticles in the bulk
fuse to total non-Abelian charge $1$ or $\psi$, respectively;
$J_{-1/4}$ is the Bessel function; $\Phi$ is the flux enclosed in the interference loop; and
$n$ is the (even) number of bulk quasiparticles inside the loop. The phase $n\pi/4$
is statistical phase due to the Abelian part of the theory.
The phase $\alpha$ is $\text{arg}({\Gamma_1}{\Gamma_2^*})$.
When the charge and neutral velocities are not equal, the current and differential conductance will oscillate at two different frequencies as seen in Figure (\ref{Conductance}), and both charge and neutral velocities can be extracted from the two different periods. The smaller period corresponding to the fast oscillations is roughly $\frac{16\pi}{e|x_1-x_2|}(1/v_n+1/v_c)^{-1}$, and the larger period corresponding to the oscillations of the envelope is roughly $\frac{16\pi}{e|x_1-x_2|}(1/v_n-1/v_c)^{-1}$.

Finite-temperature correlation functions can be obtained from the zero temperature correlation functions by a conformal transformation
from the plane to the cylinder, which amounts to
the following substitution:
\be\label{TempSub}
\frac{1}{\left(\delta+i(t\pm x/v)\right)^{1/8}}\rightarrow
\left(\frac{\pi T}{\sin\left(\pi T(\delta+i(t\pm x/v))\right)}\right)^{1/8}
\ee
We find that the general form of the current is:
\begin{equation}
{I_1}+{I_2} = \left({|\Gamma_1|^2}+{|\Gamma_2|^2}\right)\,
|V|^{-1/2}\, A\!\left(\frac{eV/4}{{k_B}T}\right)
\end{equation}
\begin{multline}
I_{12} = |\Gamma_1||\Gamma_2| \,\cos\!\left(\frac{\Phi}{4\Phi_0}+
n\frac{\pi}{4}+\alpha\right) \text{sgn}(V)
|V|^{-1/2}\,\times\\
{B_n}\!\left(\frac{e|x_1-x_2|}{4 v_c}|V|,  \frac{e|x_1-x_2|}
 {4 v_n}|V|, \frac{eV/4}{{k_B}T}\right)
\end{multline}
where $B_{2n+1}(x,y,z)=0$, and
$A(x)$ and $B_{2n}(x,y,z)$ are scaling functions which reduce
to (\ref{eqn:zero-temp-direct}) and (\ref{eqn:zero-temp-int})
in the $T=0$ limit: $A(\infty)=\text{const.}$, $B_{2n}(x,x,0)\propto x^{1/4}\, J_{-1/4}(x)$.
In the opposite limit, ${k_B}T>eV$, $A(x)\sim x^{3/2}$ as $x\rightarrow 0$,
so that the conductance due to a single point contact is
$\sim T^{-3/2}$. The explicit form of $A(x)$ is
$$
A(x)= \frac{1}{\sqrt{2}}  \, \frac{e}{4} \, \frac{\pi \sqrt{x}}{\Gamma(\frac{1}{2})}\left|\Gamma\!\left(
\frac{1}{4}+ i\frac{x}{2\pi}\right)\right|^2 \sinh(x/2)
$$

$B_{2n}(x,y,z)$ is more complicated, but
it simplifies in the limit that $(x+y)/z$ is large,
where $B_{2n}(x,y,z) \sim e^{-(x+y)/z}$.
Consequently, there is an effective dephasing length \cite{LeHur02,LeHur05,LeHur06}
\begin{equation}
L_\phi = \frac{\beta}{2\pi} \left(\frac{1/8}{v_c}+\frac{1/8}{v_n}\right)^{-1}
\end{equation}
such that
\begin{equation}
\label{eqn:visibility}
I_{12} \propto e^{-|x_1-x_2|/L_\phi}\,\cos\!\left(\frac{\Phi}{4\Phi_0}+
n\frac{\pi}{4}+\alpha\right)
\end{equation}
Interference is only visible if the interferometer is
smaller than $L_\phi$.
Equivalently, there is a characteristic temperature
scale\cite{Fidkowski07} $T^*$:
\be
k_B {T^*}=
\frac{1}{2\pi|x_1-x_2|}\left(\frac{1/8}{v_c}+\frac{1/8}{v_n}\right)^{-1}
\ee
Interference is only visible for $T<T^*$ since
(\ref{eqn:visibility}) can be rewritten as:
\begin{equation}
I_{12} \propto e^{-T/T^*} \,\cos\!\left(\frac{\Phi}{4\Phi_0}+
n\frac{\pi}{4}+\alpha\right)
\end{equation}
For fixed $v_c$, decreasing $v_n$ causes to ${T^*}$
and $L_\phi$ to decrease.
If $v_n$ becomes very small, interference will only be visible
at extremely low temperatures or for extremely
small interferometers (which, of course, suffer from other problems).
In the extreme limit, ${v_n}=0$, interference will not be visible at all.
Numerical studies \cite{Wan06} indicate that the two velocities
might be quite different, in which case, it will be
important that interferometry experiments be done
at sufficiently low temperatures. Using commonly accepted values of edge velocities (see, for instance, Ref. [\onlinecite{Wan07}]) of $v_c\approx 5\cdot 10^4 m/s$ and $v_n=0.1\,v_c$, we estimate the dephasing length $L_\Phi$ to be about $4 \mu m$ at a temperature of $10mK$. We will see below that the direction of the propagation of the neutral mode is irrelevant for these DC interference measurements. Even when the neutral modes propagate in opposition to the charge modes, as in the anti-Pfaffian state, interference can be observable, and the dephasing length is only a function of the magnitude of the velocities of the edge modes.

Figure (\ref{Conductance}) shows the differential conductance
$\partial I/\partial V$ at a temperature much lower than $T^*$
for both even and odd numbers of bulk quasiparticles. As may be seen
from this figure, the difference between even and odd numbers of
quasiparticles is still very dramatic, even for finite temperature
and different charge and neutral velocities. The even quasiparticle
differential conductance passes through zero twice at voltages
which are small enough that the odd quasiparticle differential conductance
is still appreciable (and, of course, due entirely to $I_{1}+I_{2}$).

\section{Anti-Pfaffian Edge} \label{aPf}
If one ignores Landau level mixing, then the Hamiltonian for the $\nu=5/2$ FQH system is particle-hole symmetric when there is exactly half an electron per flux quantum (ignoring the filled Landau Levels). The Pfaffian state, on the other hand, does not posses this symmetry. The particle-hole conjugate of the Pfaffian state, the anti-Pfaffian ($\overline{\rm Pf}$) \cite{LeeSS07,Levin07}, has the same energy in the absence of Landau level mixing as the Pfaffian, and should be considered a candidate for the observed $\nu=5/2$ state, even with finite Landau Level mixing.

The edge theory of the anti-Pfaffian can be considered by considering a
Pfaffian state of holes in a filled $\nu=1$ Landau level:
\begin{multline}
{\cal L} =  \frac{1}{4 \pi} {\partial_x}{\phi_1}(-i{\partial_t}+v_{1}{\partial_x}){\phi_1}
+ {\cal L}_\text{Pf}({\psi_1},{\phi_2})\\
 + \frac{1}{4\pi} 2v_{12}{\partial_x}{\phi_1}{\partial_x}{\phi_2}
 + \xi(x)\, {\psi_1}\, e^{i({\phi_1}-2{\phi_2})} + \text{h.c.} .
\end{multline}
Here, ${\cal L}_\text{Pf}({\psi_1},{\phi_2})$ is the Pfaffian edge action
(\ref{eqn:Pf-edge}) but for counter-propagating edge modes.
The coupling $v_{12}$ is short-ranged Coulomb repulsion between the
edge mode of the filled Landau level and the charged edge mode of the
Pfaffian state of holes while $\xi(x)$ is random tunneling of electrons
between the $\nu=1$ edge and the edge of the Pfaffian of holes.
For large $v_{12}$ and arbitrarily weak $\xi$ or for small $v_{12}$ and
sufficiently large $\xi$, the theory flows in the infrared to a theory
of a forward propagating bosonic charge mode and three {\it backward propagating}
neutral Majorana modes\cite{LeeSS07,Levin07}:
\be
{\cal L}_{\bar{Pf}}=\frac{2}{4\pi}\partial_x\phi_\rho\left(\partial_t+v_c\partial_x\right)\phi_\rho + \sum\limits_{a=1,2,3} i\psi_a\left(-\partial_t+v_n\partial_x\right)\psi_a
\ee
We will discuss quasiparticle tunneling in this phase of the anti-Pfaffian
edge.
The three Majorana fermions form an $SU(2)_2$ triplet, which means that the non-Abelian statistics due to this part of the theory are associated with $SU(2)_2$ Chern-Simons theory\cite{Fradkin98}.
The electron operator in this theory is $(\psi_2-i\psi_3)e^{i2\phi_\rho}$.
The charge $e/4$ quasiparticles are the primary fields $\phi_{1/2}^{\pm}e^{i\phi_\rho/2}$, where $\phi_{1/2}^{\pm}$ are the spin-$1/2$ fields of $SU(2)_2$, and can be written in terms of the Ising order and disorder fields $\sigma_a$ and $\mu_a$. The $\phi_{1/2}^{\pm}$ fields consist of linear combinations of products of 3 $\sigma_a$ or
$\mu_a$ operators, and therefore has dimension $3/16$. Consequently,
the $e/4$ quasiparticle operator in the anti-Pfaffian state has dimension $1/4$, as opposed to dimension $1/8$ in the Pfaffian case.
This difference in the scaling dimension causes the Pfaffian and anti-Pfaffian
to have different temperature and voltage dependance for transport
through point contacts which, in principle, allows one to experimentally
distinguish between the two states.
Another important difference is that in the anti-Pfaffian case,
the charge $e/2$ quasiparticle operator has the same scaling dimension
as the $e/4$ quasiparticle and its tunneling is just as relevant,
but one expects the bare tunneling element for the $e/2$ quasiparticle
to be smaller than the $e/4$ one ($|\Gamma_{e/2}|\sim |\Gamma_{e/4}|^2$).

The above discussion implies that $e/4$ quasiparticle tunneling is the dominant one also in the anti-Pfaffian case. The tunneling current calculation in the double quantum point setup proceeds in a very similar fashion to the Pfaffian case.
To lowest-order, we must compute four-quasiparticle correlation functions,
and the relevant conformal block is the one in which quasiparticle fields on both ends of a point contact should fuse the identity. In the long sample limit, we seek the projection of these correlation function on the conformal block in which quasiparticles on the same edge fuse to the identity.

$SU(2)_2$ non-Abelian statistics are similar to the Ising statistics that appear in the
Pfaffian. In the $SU(2)_2$ theory there are only $3$ particle types,
$0$,$1/2$, and $1$, with the fusion rule:
\be
\frac{1}{2} \times \frac{1}{2} = 0 + 1
\ee
which is analogous to the the fusion rule in Eq.(\ref{sigmasigma}). Hence, the enumeration of conformal blocks in $SU(2)_2$ theory is the same as in the Ising theory if we identify the operators $I$,$\sigma$ and $\psi$ with $0$, $1/2$, and $1$
operators respectively. Also, the matrix elements of the F-matrix which describes the
change of basis between different fusion channels turn out to be the same in both theories, up to a phase\cite{DasSarma07}.
An equation analogous to Eq.(\ref{LongSample}) holds for the anti-Pfaffian case also, but with different power laws since the spin $1/2$ operator has a different scaling dimension than the $\sigma$ operator:
\begin{align}
\label{LongSample-AP}
{\cal F}_I \mid_{L\rightarrow \infty} &= \tilde{a}_I {\cal G}_I\mid_{L\rightarrow_\infty} \\
&= \tilde{a}_I\prod_{\epsilon=\pm}
\left[ \delta+i\left(v_n(t-t')-\epsilon(x_j-x_k)\right)\right]^{-3/8} \nn
\end{align}

The tunneling current behavior in the anti-Pfaffian case is qualitatively
the same as but quantitatively different from the Pfaffian case.
One might worry that no interference should take place at all since the $e/4$ quasiparticle operator is made up of a bosonic part moving in one direction and a fermionic part moving in the opposite direction, and in a semiclassical picture these two parts are moving away from each other.
In fact, the sign of the neutral mode velocity makes no difference,
as may be seen by comparing (\ref{LongSample}) and
(\ref{LongSample-AP}). As a result of the product over
$\epsilon=\pm$, the sign of the neutral mode velocity
drops out of the problem. The point is that
the quantum mechanical tunneling process involves
creating a quasiparticle and a quasihole, and regardless of the chirality of the mode, one excitation will move to the left and one to the right.
We note that this breakdown of semiclassical intuition
represented by the insensitivity to the neutral mode
direction is a feature of a DC measurement. A finite frequency
measurement might be more sensitive to the difference between
the charge and neutral velocities.

At zero temperature in the anti-Pfaffian state,
\begin{equation}
I_{1}+I_{2}= \frac{1}{\sqrt{2}} \, \frac{e}{4}\, \pi \left(|\Gamma_1^2|+|\Gamma_2|^2\right)v_n^{-6/8} v_c^{-2/8} \,  \text{sgn}(V)
\end{equation}
The conductance will behave as $V^{-1}$; the differential conductance
will be sharply peaked at $V=0$ (with a peak width of
order ${k_B}T$) and vanishing elsewhere. For ${k_B}T>eV$,
the conductance varies as $T^{-1}$. In both cases, there are
quantitative differences from the Pfaffian.

Again, for an odd number of quasiparticles in the interference loop,
\begin{equation}
I_{12} = 0
\end{equation}
For an even number of bulk quasiparticles, the tunneling current will oscillate with magnetic field and voltage, similar to the Pfaffian case. Again, for charge and neutral velocities which are equal in absolute value (although opposite in sign),
$I_{12}$ can be found analytically:
\begin{multline}
\label{I_12-AP}
I_{12} = \pm \frac{1}{\sqrt{2}}\, \frac{e}{4}\, \frac{2\pi^{3/2}}{\Gamma(\frac{1}{2})}
 |\Gamma_1||\Gamma_2|
\cos\!\left(\frac{\Phi}{4\Phi_0}+n\frac{\pi}{4}+\alpha \right) \,\times\\
 \text{sgn}(V)\,
     J_{0}\!\left(\frac{e|x_1-x_2|}{4 v_c}|V|\right)
\end{multline}
Although the phase acquired in the anti-Pfaffian state
by an $e/4$ quasiparticle
in going around another $e/4$ quasiparticle
is different (in either fusion channel)
from in the Pfaffian state, the phase
acquired by an $e/4$ quasiparticle in going around
a charge $e/2$ is $\pm i$ in either state,
with the minus sign corresponding
to the presence of a neutral fermion.

A difference between the absolute values of the
neutral and charge velocities will again be evident through a beating pattern
in the differential conductance. $I_{12}$
is exponentially decaying with temperature
with characteristic scale:
\be
k_B {T^*}=
\frac{1}{2\pi |x_1-x_2|}\left(\frac{1/8}{v_c}+\frac{3/8}{v_n}\right)^{-1}
\ee
and the corresponding dephasing length is:
\be
L_\phi = \frac{\beta}{2\pi} \left(\frac{1/8}{v_c}+\frac{3/8}{v_n}\right)^{-1}.
\ee
\section{Discussion}

As we have seen from the preceding formulas,
the Pfaffian and anti-Pfaffian state have qualitatively similar
behavior in a two point-contact interferometer. In particular,
the reversal of the neutral modes in the latter state makes
little difference. However, the temperature and voltage
dependences of the backscattered current
are quantitatively different. The difference is clear
in the behavior of a single-point contact, where the
associated power laws are different, $I\sim V^{-1/2}$
in the case of the Pfaffian and $I\sim V^{0}$ in the
case of the anti-Pfaffian. However, there are also
differences in the detailed temperature
and voltage dependence of the interference contribution
to the current, as may be seen from Eqs. \ref{eqn:zero-temp-int},
\ref{I_12-AP}.

The relative insensitivity of quantum interference
effects to the difference between the charge and neutral
mode velocities runs counter to semi-classical thinking
(and shows its limitations): naively, one might think that
when a quasiparticle decays into its charged and neutral
parts, interferometry would be hopeless. Fortunately,
this is not the case, as explicit calculation shows.
This also augurs well for the suitability of either one for quantum computation along the lines of Refs.
\onlinecite{DasSarma05,Freedman06}.
The downside is that the experimental difference between the
Pfaffian and anti-Pfaffian states is muted. It can be
extracted from the behavior in an interferometer,
but it would still be useful to have a probe which is
more sensitive to the direction of the neutral modes.

\acknowledgements
We would like to thank E.\ Ardonne and L.\ Fidkowski for helpful discussions.

\bibliographystyle{prsty}

\end{document}